\documentclass[twocolumn,showpacs,preprintnumbers, prc,superscriptaddress]{revtex4}
\usepackage{epsfig}
\usepackage{graphicx}
\usepackage{epstopdf}
\usepackage{amsmath,amssymb,amsfonts}
\usepackage{array}
\usepackage{url}
\usepackage{hyperref}
\usepackage{lineno}
\usepackage{xspace}
\usepackage[usenames,dvipsnames]{color}
\newcommand{\sqsn}{\mbox{$\sqrt{s_{_{NN}}}$}\xspace}
\newcommand{\bef}{\begin{figure}}
\newcommand{\eef}{\end{figure}}
\newcommand{\bc}{\begin{center}}
\newcommand{\ec}{\end{center}}

\newcommand{\auau}{\mbox{Au$+$Au}\xspace}
\newcommand{\ptot}{\mbox{$p^{\rm incl}$}\xspace}
\newcommand{\pstop}{\mbox{$p^{\rm stop}$}\xspace}
\newcommand{\pprod}{\mbox{$p^{\rm prod}$}\xspace}
\newcommand{\pdiff}{\mbox{$p^{\rm diff}$}\xspace}
\begin{document}
\title{Disentangling stopped proton and inclusive net-proton fluctuations at RHIC}

\author{D. K. Mishra}
\email {dkmishra@rcf.rhic.bnl.gov}
\affiliation{Nuclear Physics Division, Bhabha Atomic Research Center, Mumbai 
400085, India}
\author{P. Garg}
\email {prakhar@rcf.rhic.bnl.gov}
\affiliation{Department of Physics and Astronomy, Stony Brook University, SUNY, 
Stony Brook, New York 11794-3800, USA}

\begin{abstract}
The recent results on net-proton and net-charge multiplicity fluctuations from the 
beam energy scan program at RHIC have drawn much attention to explore the critical 
point in the QCD phase diagram. Experimentally measured protons contain
contribution from various processes such as secondaries from higher mass resonance 
decay, production process, and protons from the baryon stopping. Further, these 
contributions also fluctuate from event to event and can contaminate the dynamical 
fluctuations due to the critical point. We present the 
contribution of stopped proton and produced proton fluctuations in the net-proton 
multiplicity fluctuation in \auau collisions measured by STAR experiment at RHIC. 
The produced net-proton multiplicity fluctuations using cumulants and their ratios 
are studied as a function collision energies. After removing the stopped 
proton contribution from the inclusive proton multiplicity distribution, a 
non-monotonic behavior is even more pronounced in the net-proton fluctuations around 
\sqsn = 19.6 GeV, both in $S\sigma$ and $\kappa\sigma^2$. The present study will be 
useful to understand the fluctuations originating due to critical point.

\end{abstract}

\maketitle

\section{Introduction}
\label{sec:intro} 
The recent Beam Energy Scan (BES) program at Relativistic Heavy Ion Collider 
(RHIC) is motivated to explore the structure of QCD phase diagram, such as a first 
order co-existence region or a critical 
point~\cite{Stephanov:1998dy,Stephanov:2008qz,Kitazawa:2013bta,Skokov:2012ds}. 
Lattice QCD calculations suggest that, there is a simple cross-over from 
quark-gluon-plasma (QGP) to hadronic phase at high $T$ and low $\mu_B$ 
~\cite{Aoki:2006we,Fodor:2004nz,Alford:1997zt}. Several other models predict a 
first order phase transition at large $\mu_B$ and low 
$T$~\cite{Stephanov:2004wx,Bazavov:2012vg}. Hence, there should be a point where the 
first order phase transition line ends, which is named as QCD critical end point 
(CEP)~\cite{Ejiri:2005wq,Stephanov:1999zu}.

The moments (mean $M$, variance $\sigma$, skewness $S$, and kurtosis $\kappa$) of 
the multiplicity distribution of conserved quantities such as, net-baryon, 
net-charge, and net-strangeness are related to the correlation length 
($\xi$) of the system~\cite{Stephanov:1998dy,Stephanov:2008qz}. Hence, event-by-event 
fluctuations of these conserved quantities can be used to look for signals of a 
critical point and phase transition~\cite{Koch:2005vg, 
Asakawa:2000wh,Asakawa:2009aj}. Most of the experimental efforts are made to measure 
the fluctuations in terms of moments or cumulants of above conserved quantities. 
The higher moments have stronger dependence on $\xi$, thus should be more sensitive 
to the critical fluctuations, originating due to phase transition trajectory that 
passes through a critical 
point~\cite{Stephanov:2008qz,Ejiri:2005wq,Stephanov:1999zu,Asakawa:2009aj,
Gavai:2010zn,Cheng:2008zh}. The individual moments (cumulants $C_n$, $n$ = 1, 2, 3, 
and 4) of the conserved distributions can have trivial system dependence, which can 
be minimized by taking the ratios of the moments. Further, the cumulant ratios can be 
related to the generalized susceptibilities calculated in lattice 
QCD~\cite{Bazavov:2012vg,Ejiri:2005wq,Cheng:2008zh} and other statistical model 
calculations~\cite{Karsch:2010ck,Garg:2013ata,Mishra:2016tne}.

The experiments at RHIC have reported the results on higher moments of 
net-proton~\cite{,Aggarwal:2010wy,Adamczyk:2013dal} and 
net-charge~\cite{Adare:2015aqk,Adamczyk:2014fia} multiplicity distributions, using 
the data from BES-I program at different center of mass energies (\sqsn) = 7.7, 11.5, 
19.6, 27, 39, 62.4, and 200 GeV. At lower collision energies, the STAR experiment 
observed a large deviation from the 
Poisson expectation in the net-proton fluctuation results~\cite{Adamczyk:2013dal}. 
Several studies have been performed to estimate the sources of excess fluctuations, 
such as the effect of kinematical acceptance~\cite{Garg:2013ata}, 
volume fluctuations~\cite{Skokov:2012ds}, inclusion of resonance 
decays~\cite{Mishra:2016qyj,Nahrgang:2014fza,Begun:2006jf}, exact (local) charge 
conservation~\cite{Bzdak:2012an,Schuster:2009jv,Netrakanti:2014mta} or repulsive 
van-der-Walls forces among hadrons~\cite{Fu:2013gga,Bhattacharyya:2015zka}. The 
effect of baryon stopping is predominant at lower collision energies, where a 
possible QCD critical point is expected to be observed. The proton and anti-proton 
pair-production is negligible at lower \sqsn, hence most of the measured protons 
are originated from the beam particles. Therefore, the dynamics of baryon stopping 
is another source of fluctuation, which can distort the signal for the critical 
point and phase-transition~\cite{Bzdak:2016jxo}. However, it is important to estimate 
the fluctuations due to baryon stopping and initial state participant 
fluctuations~\cite{Thakur:2016znw}.

\begin{figure*}[ht]
\bc
\includegraphics[width=1.0\textwidth]{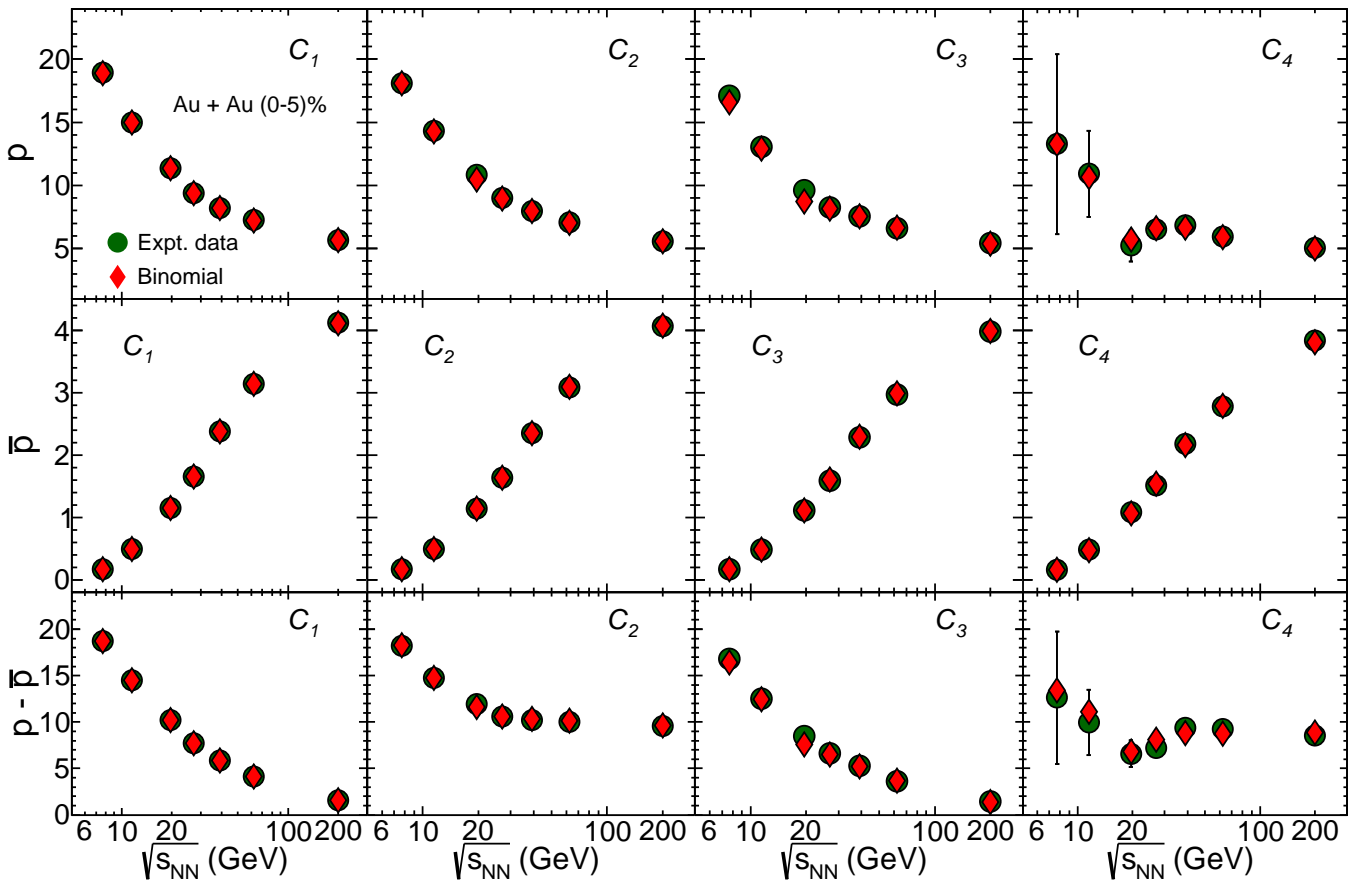}
\caption{The efficiency corrected individual cumulants of inclusive protons ($p$), 
anti-protons ($\bar p$), and net-protons ($p-\bar p$) multiplicity as a function of 
\sqsn in most central (0\%--5\%) \auau collisions. The experimental cumulants 
from~\cite{Adamczyk:2013dal,stardata} are compared with the same obtained from 
a Binomial assumption.}
\label{fig:cumu_ex_bin}
\ec
\end{figure*}

The experimentally observed baryon stopping could be used as a direct tool to 
explore the QCD phase transition~\cite{Ivanov:2010cu}. The stopping in nuclear 
collision can be estimated from the rapidity loss occurred by the baryons. 
Hence, baryon stopping is inferred through the net-baryon rapidity ($y$)
distribution~\cite{Ivanov:2010cu}. The measured net-baryon distribution retains the 
information about the energy loss and allows to determine the degree of nuclear 
stopping. The baryon stopping can be quantified by the reduced curvature of the 
net-proton rapidity distribution at mid-rapidity. Since most of the experiment do not 
measure neutrons within the same kinematical acceptance as protons, hence baryon 
stopping can be accessible through the measurement of protons and anti-protons. 
Further, the measured protons have contributions from both stopping and produced 
protons. At AGS energies, the number of produced anti-proton is very small, the 
net-baryon rapidity distribution is similar to the proton 
distribution~\cite{Back:2000ru,Ahle:1999in,Barrette:1999ry}. At SPS energies, the 
net-proton rapidity distribution shows a double hump structure with a dip around $y$ 
= 0. This suggests that, the reaction at SPS is beginning to be transparent, only 
small fraction of original baryons are found at the mid-rapidity and the hump 
structures reflect the rapidity distributions of the produced protons after the 
collisions~\cite{Anticic:2003ux}. At RHIC, the rapidity distribution for net-protons 
at \sqsn = 200 GeV is different from those at lower energies indicating a different 
system formed around the mid-rapidity~\cite{Bearden:2003hx}. The experimental data 
shows that collisions at such high energy exhibit a high degree of transparency. 
Baryon transparency, which is inverse of baryon stopping, increases with the 
collision energy. Since baryon stopping plays a major role at lower \sqsn, which 
almost vanishes at higher collision energies, it is important to disentangle the 
contribution of stopped baryons and the produced baryons in order to 
understand the QCD phase structure and critical point.

\begin{figure*}[ht]
\bc
\includegraphics[width=1.0\textwidth]{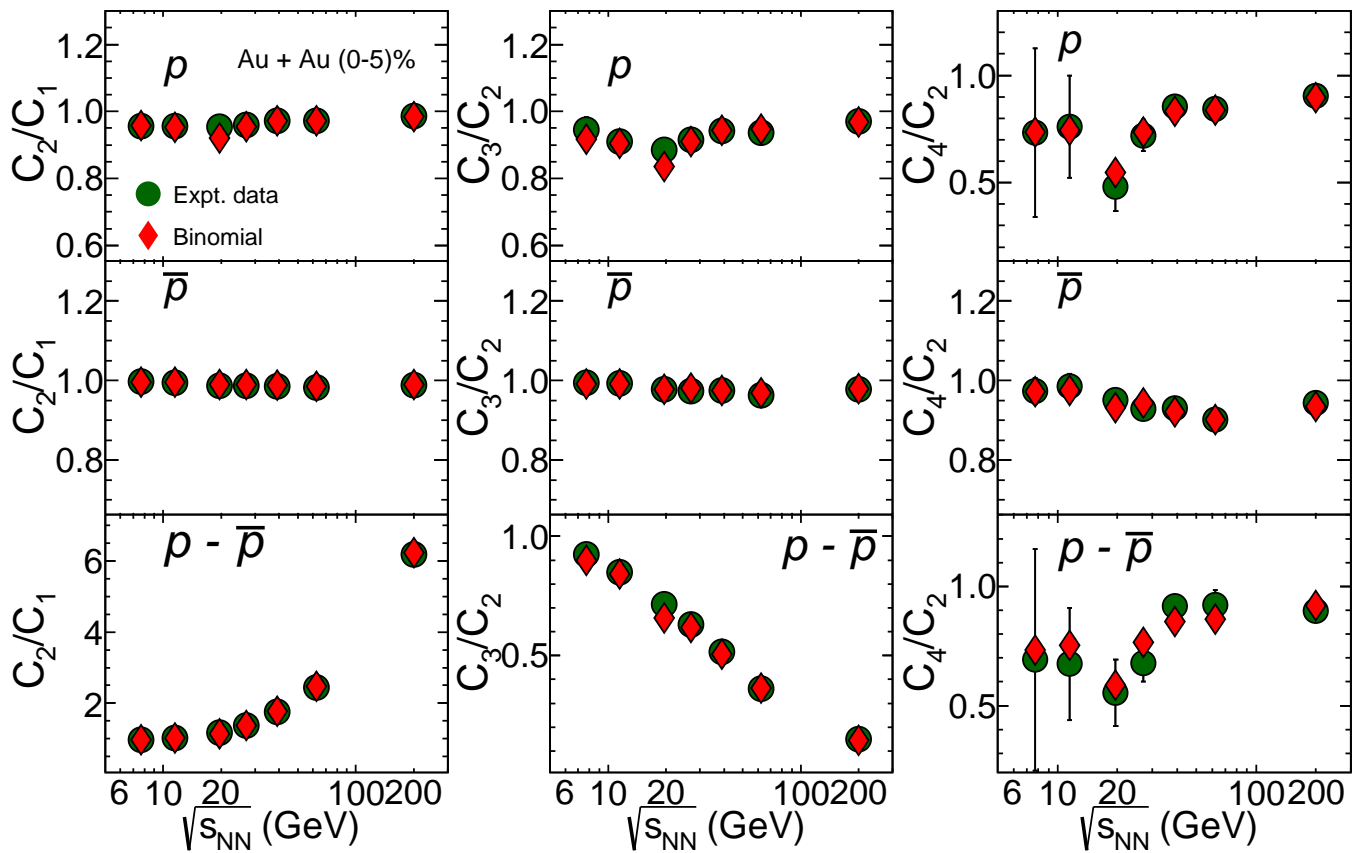}
\caption{Collision energy dependence of efficiency corrected cumulant ratios 
($C_2/C_1$, $C_3/C_2$, and $C_4/C_2$) of $p$, $\bar p$, and $p-\bar p$ as a function 
of \sqsn in most central (0\%--5\%) \auau collisions. The experimental cumulant 
ratios from~\cite{Adamczyk:2013dal,stardata} are compared with the same obtained 
from a Binomial assumption. }

\label{fig:cumu_r_ex_bin} 
\ec
\end{figure*}


The present work is motivated to study the net-proton multiplicity fluctuation, 
which is related to the particle production mechanism. Experimentally measured 
proton (anti-proton) multiplicity distributions contain the 
contributions from nuclear stopping, resonance decay, and also directly produced from 
the collisions. A detailed studies of resonance contribution to net-proton 
fluctuations have been reported in Refs.~\cite{Mishra:2016qyj,Nahrgang:2014fza}. The 
previous net-proton fluctuation studies~\cite{Aggarwal:2010wy,Adamczyk:2013dal} by 
STAR experiment have been performed by using higher moments with inclusive proton 
(anti-proton) multiplicity distributions. However, after subtracting the stopping 
contribution from the inclusive proton multiplicity distribution, it may have large 
effect in the correlation of proton and anti-proton multiplicities. This can affect 
the higher moments of the net-proton multiplicity 
fluctuations~\cite{Mishra:2015ueh}. It will be interesting to look for higher moments 
of net-proton fluctuations after removing the stopped proton contribution, which is 
dominant at lower \sqsn.

The paper is organized as follows: In the following section, we discuss the method 
used for the present study. In Sec.~\ref{sec:results}, the results of individual 
contribution to the proton fluctuations and net-proton fluctuations are shown. 
Finally, we summarize our work and discuss it's implications in 
Sec.~\ref{sec:summary}.

\begin{table*}[hbt]
 \caption{Mean values of inclusive proton, anti-proton, and stopped proton 
distributions for most central (0$\%$--5$\%$) \auau collisions at various \sqsn 
measured by STAR experiment~\cite{Adamczyk:2013dal,Thakur:2016znw,stardata}.}
\begin {tabular}{cccccccc}
\hline
\hline
\sqsn (GeV) & 7.7 & 11.5 & 19.6 & 27 & 39 & 62.4 & 200\\
\hline
Incl. proton&        $18.918\pm0.009$ & $15.005\pm0.006$ & $11.375\pm0.003$ & 
$9.390\pm0.002$ & $8.221\pm0.001$ & $7.254\pm0.002$ & $5.664\pm0.001$\\
anti-proton& $0.165\pm0.001$  & $0.490\pm0.001$  & $1.150\pm0.001$  & 
$1.652\pm0.001$ 
& $2.379\pm0.001$ & $3.135\pm0.001$ & $4.116\pm0.001$\\
stopped proton& $17.21 \pm 0.86$ & $12.89 \pm 0.86$ & $9.73 \pm 0.80$ &
$7.61 \pm 0.73$ & $5.78 \pm 0.65$ & $3.78 \pm 0.54$ & $1.54 \pm 0.33$\\
\hline
\hline
\end{tabular}    
 \label{tbl:mean}
\end{table*}

\section{Method}
\label{sec:method} 
The STAR experiment at RHIC reported the results on the cumulants and their ratios of 
the net-protons multiplicity distributions at various collision 
energies~\cite{Adamczyk:2013dal} by measuring the number of proton and anti-proton 
multiplicities on an event-by-event 
basis. As mentioned in the previous section, the measured protons (\ptot) have 
contributions from the stopped protons (\pstop) and produced protons (\pprod). In 
the present work, we make an attempt to estimate the contributions of stopped and 
produced proton to the net-proton fluctuation results. 

A Monte-Carlo approach is adopted by taking two independent distributions of proton 
(\ptot) and anti-proton ($\bar p$) multiplicities as Binomial distribution. The 
mean values used in the Binomial distribution of \ptot and $\bar p$ for ($0\%-5\%$) 
centrality in \auau collisions at different \sqsn are taken from 
Refs.~\cite{Adamczyk:2013dal,stardata}, also shown in Table~\ref{tbl:mean}. The 
net-proton multiplicity \pdiff (= $\ptot - \bar p$) distribution is then constructed 
from independently produced \ptot and $\bar p$ distributions. For the proof of 
principle, we calculate the individual cumulants of measured proton, anti-proton 
and net-proton multiplicities, which are distributed binomially and compare them 
with the experimentally measured cumulants. Also, by this construction, we get access 
to the efficiency corrected distributions of protons and anti-protons.

\bef[ht]
\bc
\includegraphics[width=0.5\textwidth]{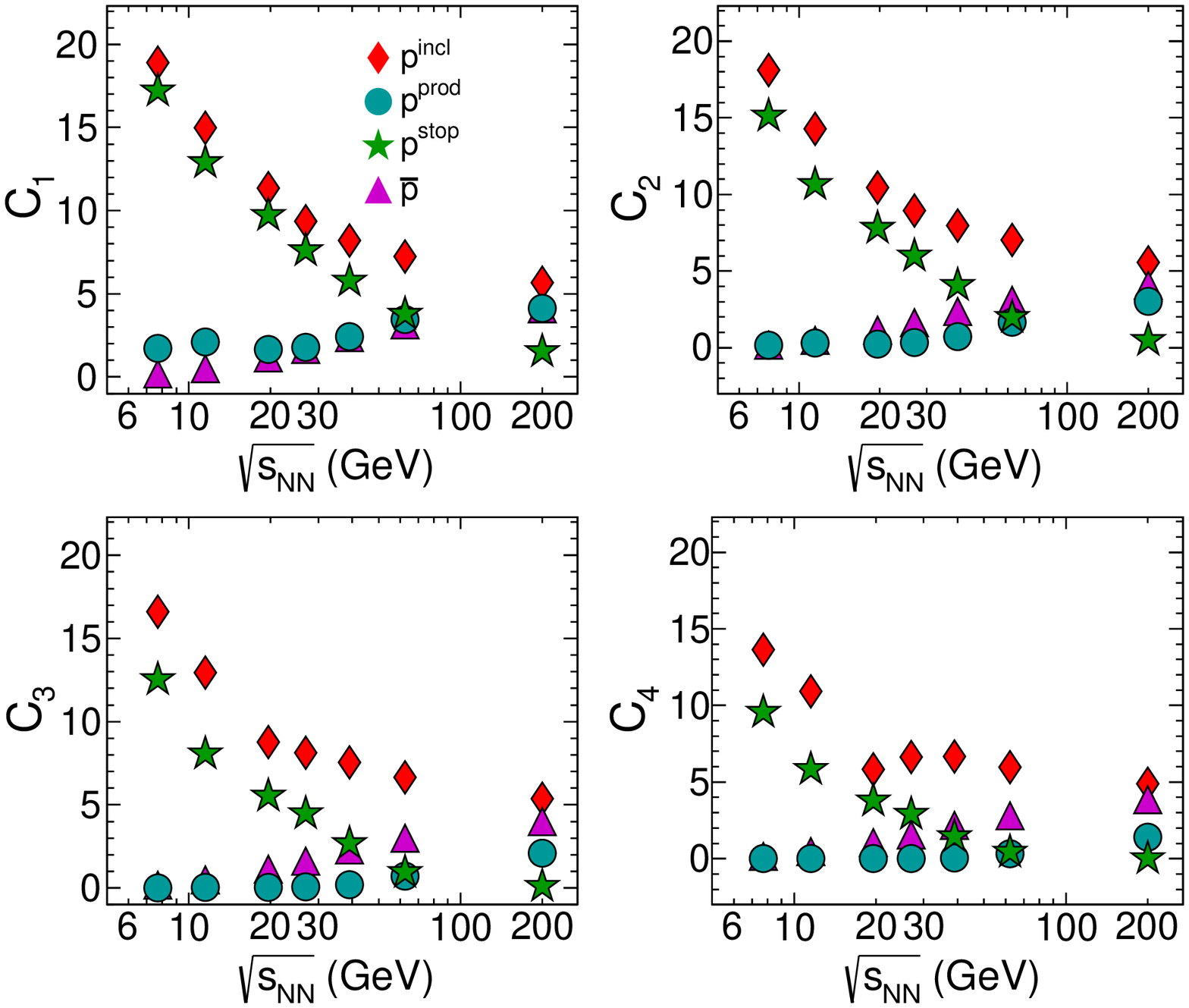}
\caption{Collision energy dependence of cumulants $C_1$, $C_2$, $C_3$, and 
$C_4$ of inclusive protons (\ptot), produced protons (\pprod), stopped protons 
(\pstop), and anti-protons in most central (0\%--5\%) \auau collisions obtained from 
Binomial distributions.}
\label{fig:cumu_indi}
\ec
\eef
\bef[t]
\bc
\includegraphics[width=0.4\textwidth]{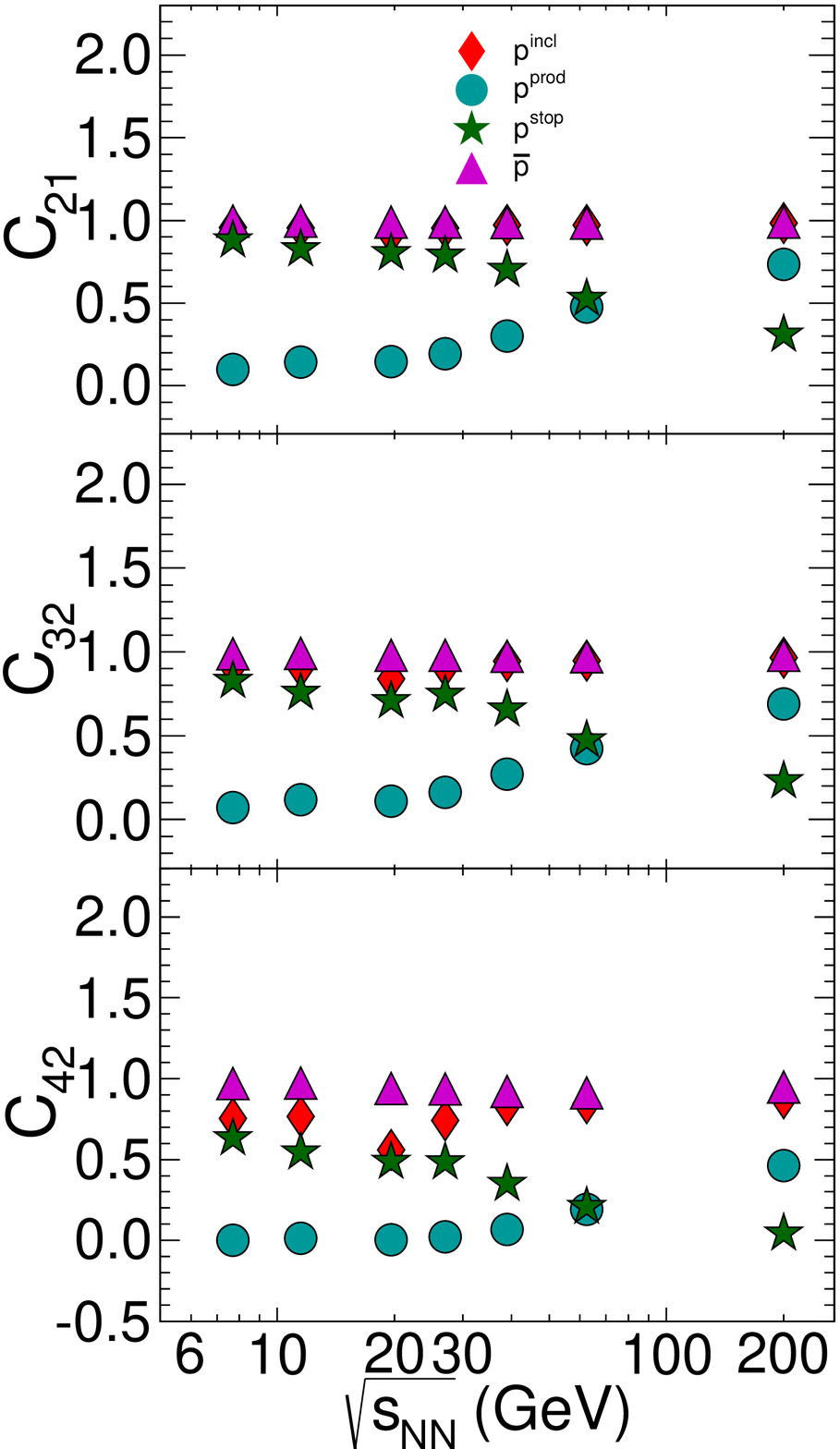}
\caption{Collision energy dependence of cumulant ratios ($C_{2}/C_{1}$, 
$C_{3}/C_{2}$, $C_{4}/C_{2}$, and $C_{3}/C_{1}$) of \ptot, \pprod, \pstop, 
and anti-protons in most central (0\%--5\%) \auau collisions obtained from Binomial 
distributions.}
\label{fig:cumu_ratios_indi}
\ec
\eef

Figure~\ref{fig:cumu_ex_bin} shows comparison of experimentally measured individual 
cumulants for proton, anti-proton, and net-proton multiplicity distributions for 
($0\%-5\%$) centrality in \auau collisions with the same obtained from Binomial 
expectation. It is to be noted that, the experimentally measured cumulants are 
corrected for reconstruction efficiency and finite centrality bin width correction. 
We tried simultaneously to reproduce all the cumulants ($C_1$, $C_2$, $C_3$, and 
$C_4$) at each \sqsn using the binomially distributed \ptot and $\bar p$. Also the 
cumulants for net-proton multiplicity distributions are well reproduced at each 
collision energy. Figure~\ref{fig:cumu_r_ex_bin} shows the cumulant ratios 
($C_2/C_1$, $C_3/C_2$, and $C_4/C_2$) of $p$, $\bar p$, and net-proton multiplicities 
at different \sqsn, which are calculated from the individual cumulants shown in 
Fig.~\ref{fig:cumu_ex_bin}. Hence, it is established that, using the Binomial 
distribution, the cumulant ratios are also well reproduced at each 
collision energy for $p$, $\bar p$, and net-proton cases. The maximum 
deviation, which was observed in experimental data for $C_4/C_2$ values of proton and 
net-proton cases at \sqsn = 19.6 GeV are also well reproduced. In the remaining 
study of this paper, we have used the above Binomial distributions which describe all 
the efficiency corrected cumulants and their ratios of the experimental net-proton 
data. In the real experimental situation, it is difficult to correct the efficiency 
of the proton, anti-proton and net-proton multiplicity distributions on an 
event-by-event basis. Hence, using the above method, one can obtain the 
efficiency corrected multiplicity distributions.

The widely discussed net-proton fluctuation study by STAR experiment has 
contribution from inclusive proton and produced anti-protons. Further, the inclusive 
protons (\ptot = \pstop~$+$~\pprod) have contributions from both stopped protons and 
produced protons. Whereas stopping has no contribution to the anti-proton 
production, they come from from resonance decay and produced particles 
in the collisions. The mean number of stopped protons \pstop at various BES energies 
have been estimated using net-proton rapidity distributions for the most central 
collisions measured by different experiments~\cite{Thakur:2016znw}. The reported 
\pstop values are in the same kinematical acceptance (transverse momentum 0.4 $\le 
p_{\rm T}~(\rm {GeV}/c) \le$ 0.8 and pseudo-rapidity $|\eta| <$ 0.5) as the STAR 
experimental data, therefore they are used directly in the present study. 

Experimentally, it is not trivial to tag a measured proton originating from stopping 
or production, hence, the correction for the stopped protons to the 
net-proton multiplicity distribution can not be applied to the experimental 
measurement. The fraction of stopped and produced proton contributions in the 
inclusive proton distribution are estimated by taking the mean number of 
\ptot~\cite{stardata} and \pstop~\cite{Thakur:2016znw}. The stopped and produced 
proton fractions are calculated as: $f^{\rm stop}= (C_1$ of \pstop/ $C_1$ of \ptot)
and $f^{\rm prod}= [(C_1$ of \ptot-- $C_1$ of \pstop)/$C_1$ of \ptot], respectively.
The stopped proton and produced proton distributions are constructed on 
event-by-event basis from the inclusive proton distribution weighted by 
corresponding $f^{\rm stop}$ and $f^{\rm prod}$ fractions, respectively. Hence, the 
mean of the original \ptot distribution change to the corresponding mean values of 
\pstop and \pprod, without modifying the shape of the distribution. 
The resulting \pstop and \pprod distributions also remain Binomial. These factorized 
multiplicity distributions along with the anti-proton multiplicity distributions are 
further used for the study of net-proton fluctuation at various collision energies.

\bef[t]
\bc
\includegraphics[width=0.5\textwidth]{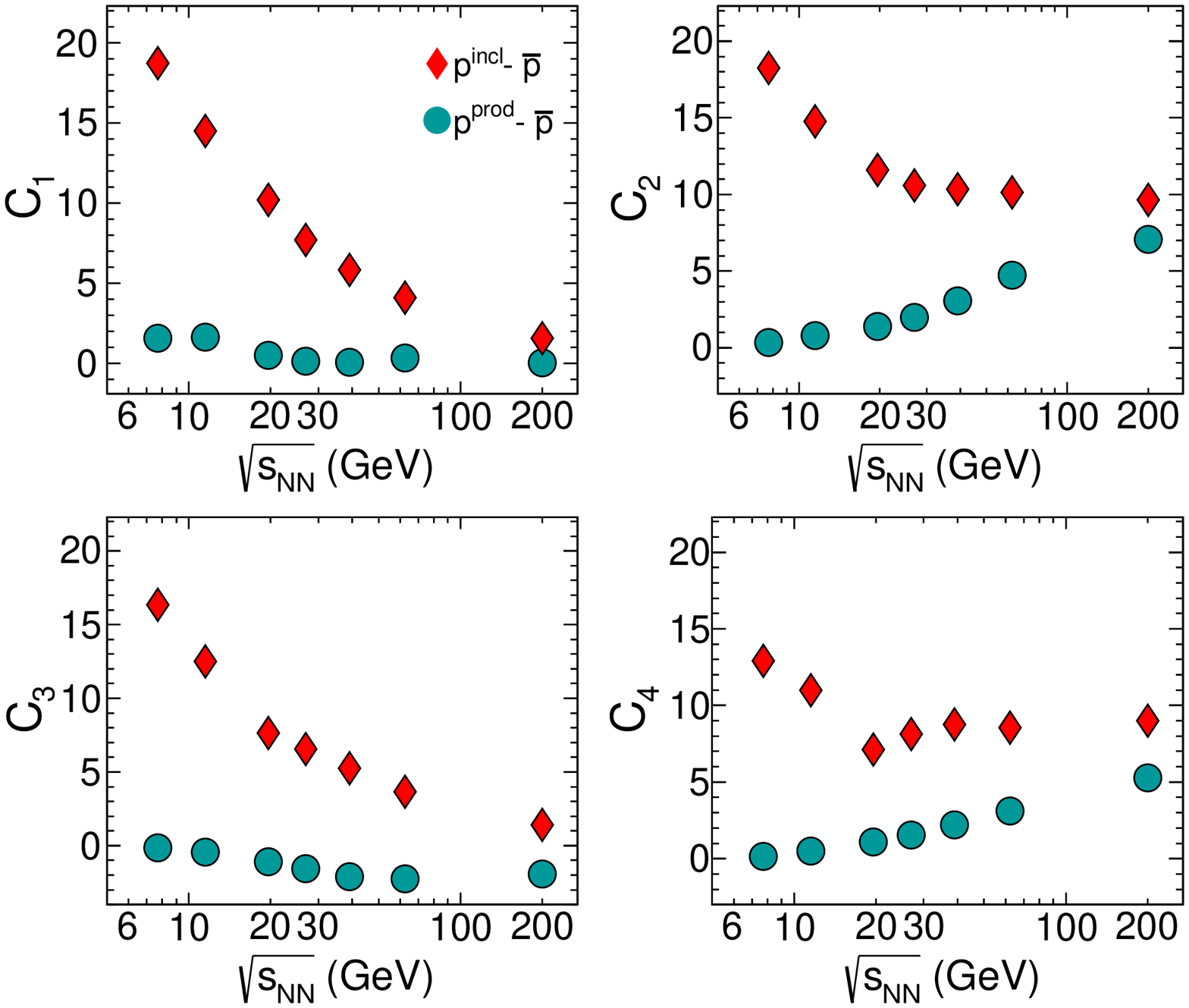}
\caption{Variation of cumulants of net-proton multiplicity distributions as a 
function of \sqsn in most central (0\%--5\%) \auau collisions obtained from Binomial 
assumption. The net-proton multiplicity distributions are calculated by taking 
\ptot -- $\bar p$ and \pprod -- $\bar p$ distributions, assuming individual \ptot and 
$\bar p$ distributions as Binomial.}
\label{fig:cumu_indi_diff}     
\ec
\eef
\bef[t]
\bc
\includegraphics[width=0.45\textwidth]{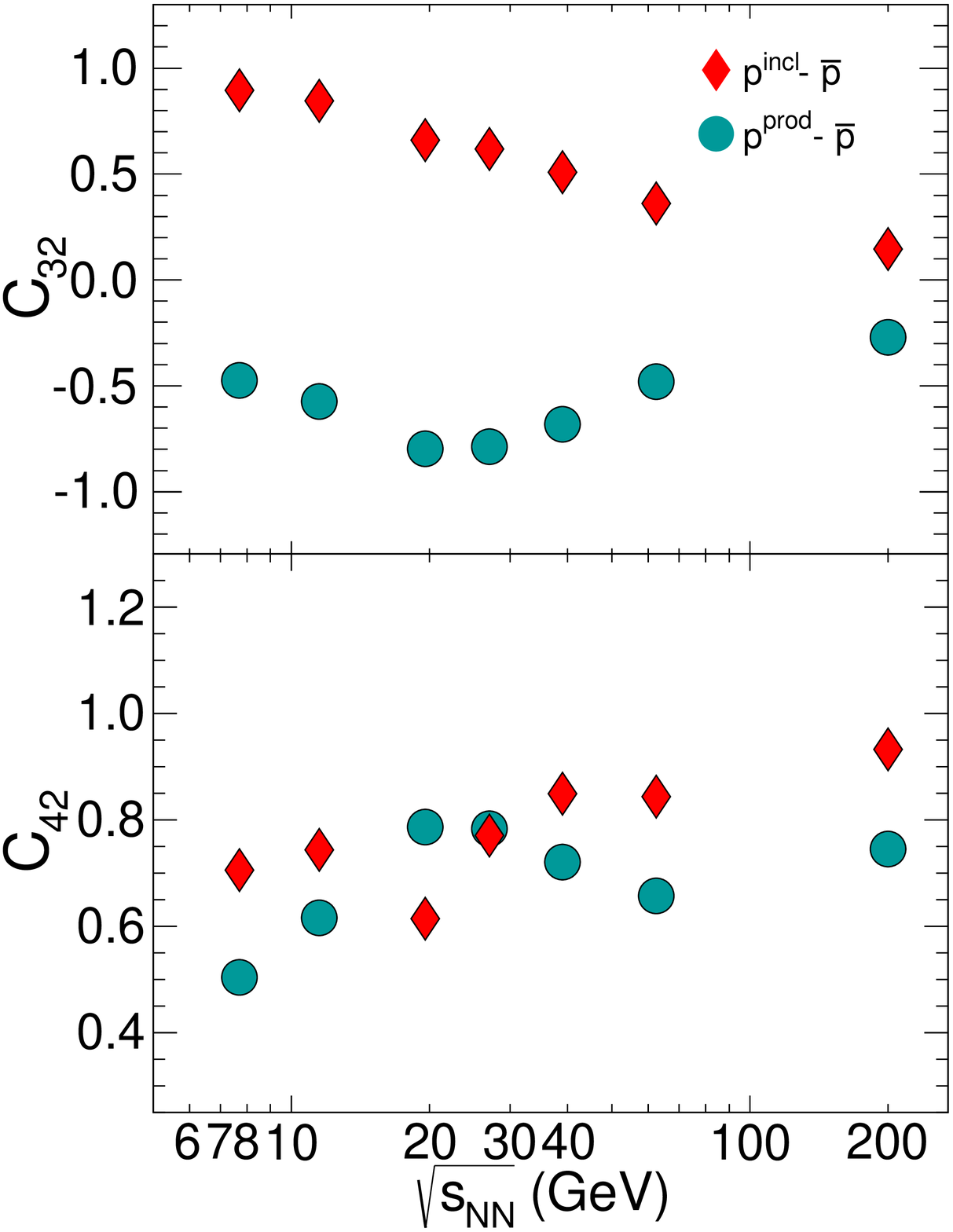}
\caption{Variation of cumulant ratios $C_{3}/C_{2}$ and $C_{4}/C_{2}$ of net-proton 
distributions as a function of \sqsn in most central (0--5\%) \auau collisions 
obtained from Binomial assumption. The net-proton distributions are calculated by 
taking \ptot -- $\bar p$ and \pprod -- $\bar p$ distributions, assuming individual 
\ptot and $\bar p$ distributions as Binomial.}
\label{fig:cumu_ratios_indi_diff}
\ec
\eef

\section{Results and Discussions}
\label{sec:results} 
The cumulants are calculated from the reconstructed \ptot, \pstop, and \pprod 
distributions. Figure~\ref{fig:cumu_indi} shows the individual 
cumulants ($C_1$, $C_2$, $C_3$, and $C_4$) of the inclusive proton, produced 
proton, stopped proton, and anti-proton multiplicity distributions as a function of 
center of mass energy obtained from the Binomial distributions as described in 
Sec.~\ref{sec:method}. All the cumulants of \ptot decrease with increasing \sqsn and 
the $\bar p$ cumulants show opposite trend. At lower energies (below \sqsn = 39 
GeV), the cumulants of \pstop and \ptot follow each other closely. This indicates 
that at lower energies, the dominant contribution to the \ptot fluctuation come from 
the \pstop. The cumulants of produced protons and the anti-protons follow close to 
each other, which further strengthen 
the argument. This behavior is consistent with the picture of evolution of 
baryon stopping and nuclear transparency with collision energies.
Figure~\ref{fig:cumu_ratios_indi} shows the 
cumulant ratios ($C_2/C_1$, $C_3/C_2$, and $C_4/C_2$) for the above mentioned (\ptot, 
\pstop, \pprod, and $\bar p$) distributions as a function of \sqsn. As observed in 
experimental data, the cumulant ratios of \ptot and $\bar p$ are similar except for 
the $C_4/C_2$ ratios at lower energies (\sqsn $<$ 39 GeV). The cumulant ratios for 
\pstop and \pprod show opposite trend as a function of collision energy. The baryon 
stopping contribution decreases with increasing collision energies, whereas the 
produced proton contribution increases. The individual cumulants and their ratios 
for \pprod and \pstop match at \sqsn = 62.4 GeV, indicating the contribution from the 
stopping and the produced protons is almost similar to the inclusive proton 
production. Further, this may also indicate that, there is equal contribution to the 
nuclear stopping and transparency in rapidity space at \sqsn = 62.4 
GeV~\cite{Wong:2000cp} in STAR acceptance.

The net-proton \pdiff cumulants are obtained by taking the number of \ptot and $\bar 
p$ multiplicity distributions on an event-by-event basis. We have also estimated the 
net-proton cumulants of the produced particles by taking the \pprod and $\bar p$ 
multiplicity distributions. Figure~\ref{fig:cumu_indi_diff} shows the individual 
cumulants of net-proton multiplicity distribution as a function of \sqsn. The 
cumulants of the difference distribution obtained from \ptot -- $\bar p$ decrease as 
a function of collision energy. As can be observed in Fig.\ref{fig:cumu_indi}, after 
subtracting the stopping contributions from the \ptot distribution, the average 
number of produced protons and anti-protons are similar, which gives a smaller value 
of $C_1$ in \pprod -- $\bar p$ distribution and it is independent of collision 
energy. The cumulants ($C_1$ and $C_3$) of the difference distributions obtained 
from \pprod -- $\bar p$ multiplicities show small energy dependence, whereas $C_2$ 
and $C_4$ cumulants systematically increase with \sqsn.  
Figure~\ref{fig:cumu_ratios_indi_diff} shows the energy dependence of ratios of 
cumulants ($C_3/C_2$ and $C_4/C_2$) for 
net-proton \pdiff distributions obtained from \ptot, \pprod, and $\bar p$ 
multiplicity distributions. The $C_3/C_2$ ratios of net-proton distributions by 
taking inclusive protons and anti-protons decrease with collision energies, similar 
observation is also reported by STAR experiment~\cite{Adamczyk:2013dal}. Further, 
the $C_3/C_2$ ratios from the produced protons and anti-protons as a function of 
\sqsn, are found to be minimum around 19.6 GeV. The $C_4/C_2$ ($= \kappa\sigma^2$) 
ratios from \ptot -- $\bar p$ shows deviation from the Poisson expectation for lower 
collision energies and maximum deviation is observed at 19.6 GeV, which is also 
observed in the experimental data. The $C_4/C_2$ ratios obtained from \pprod -- $\bar 
p$ increase with \sqsn up to 19.6 GeV and remain constant there after. 
While taking the inclusive proton with $\bar p$, a minimum is observed at 19.6 GeV 
for $C_4/C_2$ ratios, whereas by taking the produced proton and 
anti-proton a maximum is observed at the same energy. After removing the stopping 
contribution from the inclusive proton distribution, the $C_3/C_2$ and $C_4/C_2$ 
ratios of the net-proton (\pdiff) multiplicity distribution shows non-monotonic 
behavior at \sqsn = 19.6 GeV, which may indicate the presence of a critical 
point or phase-transition in heavy-ion collisions. 

\section{Summary}
\label{sec:summary} 
 To summarize the present work, we demonstrate a method to disentangle the 
contribution of the stopped protons and produced protons from the inclusive proton 
fluctuations in the heavy-ion collisions. The \sqsn dependence of the efficiency 
corrected cumulants and their ratios for proton, anti-proton, and net-proton 
multiplicities are compared with the same obtained from a Binomial distributions. The 
individual cumulants and their ratios for \ptot, \pprod, \pstop, and $\bar p$ are 
studied as a function of collision energy. At lower collision energies, the \ptot 
fluctuations are dominated by \pstop and at higher energies they are 
dominated by produced proton fluctuations. The contribution from stopping and 
produced protons to the inclusive proton fluctuations at \sqsn = 
62.4 GeV are similar. The net-proton fluctuations calculated using \ptot -- $\bar p$ 
and  \pprod -- $\bar p$ distributions are studied as a function of \sqsn. All the 
individual cumulants 
from \ptot -- $\bar p$ decrease with \sqsn. The $C_1$ and $C_3$ of \pprod -- 
$\bar p$ remain constant as a function of \sqsn. The $C_2$ and $C_4$ values 
increase with collision energies. The cumulants ratios $C_3/C_2$ and $C_4/C_2$ are 
studied as a function of collision energies for the above two different net-proton 
cases. There is a minimum observed in $C_4/C_2$ ratio at \sqsn = 19.6 GeV for 
net-proton distribution using the inclusive proton. Whereas, after 
removing the stopping contribution from the inclusive proton distribution, the 
$C_3/C_2$ and $C_4/C_2$ ratios of the net-proton multiplicity distribution 
shows non-monotonic behavior around \sqsn = 19.6 GeV. This may indicate the 
presence of a critical point or a phase-transition in high-energy heavy-ion 
collisions. This is a first such attempt to separate the stopped and produced proton 
fluctuations from the net-proton fluctuations.


\end{document}